\documentclass[aps,prd, nofootinbib,preprintnumbers]{revtex4}

\usepackage{graphicx}
\usepackage{amsmath,amssymb}
\usepackage{epsfig}

\begin{document}
\title{Boosted black string bombs}

\author{Jo\~ao G. Rosa}
\affiliation{ SUPA, School of Physics and Astronomy, University of Edinburgh, Edinburgh, EH9 3JZ United Kingdom,}
\email{joao.rosa@ed.ac.uk}

%\author{Jo\~ao G. Rosa\\ SUPA, School of Physics and Astronomy,\\ University of Edinburgh, Edinburgh, EH9 3JZ, UK\\ Email: \email{joao.rosa@ed.ac.uk}}
%\email{joao.rosa@ed.ac.uk}
% \affiliation{SUPA, School of Physics and Astronomy, University of Edinburgh, Edinburgh, EH9 3JZ, UK}

\begin{abstract}
We study the formation of superradiant bound states for massive scalar fields in five-dimensional rotating black string geometries with a non-vanishing Kaluza-Klein momentum along the compact direction. Even though all Kaluza-Klein modes may form bound states in this geometry, in realistic extra-dimensional models and astrophysical black holes only the zero-mode is sufficiently light for superradiant instabilities to develop, provided the field has a small but non-vanishing mass, as for example for axion-like particles. We use analytical and numerical methods to show that, although the Kaluza-Klein momentum decreases the upper bound on the field mass for an instability to develop, it may enhance its maximum growth rate by more than 50\%, thus boosting the black hole bomb mechanism. We discuss the possible observational consequences of this result and its potential as an astrophysical probe of non-trivial extra-dimensional compactifications. 
\end{abstract}

\date{\today} 

%\preprint{Edinburgh 2012/18}

\maketitle
%%%%%%%%%%%%%%%%%%%%%%%%%%%%%%%%%%%%%%%%%%%%%%%%%%%%%%%%%%%%%%%%%%%%%%%%%%%%%%%%%%%%%%%%%%%%%%%%%%%%%%%%%%%%%%%%%%%%%%%%%%%%%%%%%%%%%%%%%%%%%%%%%%%%%%%%%%%%%%%%%%%%%%%%%%%%%%%%%%%%%%%%%%%%%%%%%%%%%%%%%%%%%%%%%%%%%%%%%%%%%%%%%%%%%%%%%%%%%%%%%%%%%%%%%%%%%%%%%%%%%%%%%%%%%%%%%%%%%%%%%%%%%%%%%%%%%%%%%%%%%%%%%%%%%%%%%%%%%%%%%%%%%%%%%%%%%%%%%%%%%%%%%%%%%%%%%%%%%%%%%%%%%%%%%%%%%%%%%%%%%%%%%%%%%%%%%%%%%%%%%%%%%%%%%%%%%%%%%%%%%%%%%%%%%%%%%%%%%%%%%%%%%%%%%%%%%%%%%%%%%%%%%%%%%%%%%%%%%%%%%%%%%%%%%%%%%%%%%%%%%%%%%%%%%%%%%%%%%%%%%%%%%%%%%%%%

\section{Introduction}

Superradiant instabilities of Kerr black holes have been shown to be an important aspect not only in analyzing the stability of rotating black holes \cite{Zeldovich:1971mw, Teukolsky, Press:1973zz, Damour-1976, Zouros:1979iw, Detweiler:1980uk, Putten, Aguirre:1999zn, Furuhashi:2004jk,  Cardoso:2004nk, Cardoso:2004zz, Cardoso:2005vk, Dias:2006zv, KZ:2006, Dolan:2007mj, Rosa, Cardoso:2011xi, Li:2012rx} but also in using astrophysical observations to probe high energy physics. These instabilities arise from the wave analogue of the Penrose process \cite{Penrose:1969pc, Christodoulou:1970wf}, in which waves of frequency $\omega<m\Omega$, where $\Omega$ is the angular velocity of the horizon and $m$ the azimuthal angular quantum number, are amplified when scattering off the ergoregion of a rotating black hole. When the field is massive, such waves may form bound states around the black hole that are continuously amplified by multiple scatterings, leading to an exponential growth rate of the associated particle number in the so-called black hole bomb effect \cite{Press:1973zz}. In particular, it has been pointed out in \cite{Arvanitaki-JMR} that many of the axion-like fields arising in string theory compactifications have a range of extremely small non-perturbative masses, $\mu\lesssim 10^{-10}$ eV, such that they may form superradiant bound states around astrophysical black holes, leading to the formation of axion clouds with a rich associated phenomenology \cite{Arvanitaki}. The superradiant growth of particle number in a given bound state is, due to Pauli blocking, an exclusive property of bosonic fields and recently it has been explicitly shown to occur for massive vector fields \cite{Pani:2012fu}. The associated growth rate is larger than for massive scalar fields, as conjectured in \cite{Rosa:2011my}, so that the black hole bomb effect can also be used to place stringent bounds on the mass of the photon and as a probe of ultra-light hidden photons in string compactifications and related scenarios.

The intrinsic higher-dimensional nature of string theory implies, however, that one should analyze the formation of superradiant bound states in the context of higher-dimensional black holes. In higher-dimensions, uniqueness theorems as for the four-dimensional Kerr-Newman family seem to be absent, and in fact a number of new vacuum solutions with different topologies have already been found. In five-dimensions, for example, there are not only solutions with horizon topology $S^3$ - the non-rotating Schwarzschild-Tangherlini \cite{Tangherlini:1963bw} and the Myers-Perry black hole \cite{Myers:1986un} - but also the Emparan-Reall black ring with $S^1\times S^2$ topology \cite{Emparan:2001wn} and extended geometries known as black strings and black branes \cite{Horowitz:1991cd}.

Superradiant instabilities have been studies in the context of higher-dimensional black strings and branes of the form $\mathrm{Kerr}_d\times \mathcal{M}_n$ in $D$-dimensional spacetimes, such that $D=d+n$ and $\mathcal{M}_n$ is a compact manifold \cite{Cardoso:2004zz, Cardoso:2005vk}, and also for boosted black strings yielding the large radius limit of doubly spinning black rings in five dimensions \cite{Dias:2006zv}. These studies have focused on massless bulk scalar fields which upon compactification give a tower of Kaluza-Klein (KK) modes, having generically shown that the massive KK-modes may only become bound to the black hole for $d=4$, although superradiant scattering may occur in general. This is related to the fact that the gravitational potential produced by a localized massive body in $d$-dimensions falls like $r^{3-d}$, whereas the repulsive angular momentum barrier present for all superradiant modes ($l\geq1$, $m\geq1$) always decreases as $r^{-2}$, so that only for $d\leq4$ may a potential well form around the black hole.

These results imply that superradiant instabilities may only be relevant for four-dimensional Kerr black holes which are extended into the extra-dimensions, forming black strings or branes. Such geometries should naturally result from the gravitational collapse of matter in $\mathbb{R}^4\times \mathcal{M}_n$, as originally argued in \cite{Horowitz:1991cd}. Even for brane-localized matter, one expects astrophysical black holes, with a horizon radius much larger than the size of the extra-dimensions, to extend into the higher-dimensional volume both for large \cite{Argyres:1998qn} and warped \cite{Chamblin:1999by} geometries, although in the latter case a deformed {\it black cigar} rather than a flat black string will more likely result from gravitational collapse. In particular, as shown in \cite{Chamblin:1999by}, localized black holes do not consistently intersect a vacuum domain wall, and so cannot be the endpoint of the gravitational collapse of brane-bound matter fields.

The above mentioned analyses have focused on massless bulk fields, with the effective mass of each mode corresponding to its KK-momentum in the compactified dimensions. However, superradiant instabilities only occur for particle masses $\mu r_+\lesssim 1$, where $r_+$ is the black hole horizon radius. In flat compact extra-dimensions, KK-modes have masses $\mu_n=n/R$, where $R$ is the compactification radius, so that unstable KK-bound states can only occur for extra-dimensions of astrophysical scales, $R\gtrsim r_+$, which is clearly excluded. While this is also the case for more generic compact geometries with a KK-mass gap, a possible exception is the Randall-Sundrum II scenario, where the extra-dimension is infinitely long and there is a continuum of graviton KK-modes, with the graviton zero-mode localized near the single brane where the Standard Model fields are confined \cite{Randall:1999vf}. Although it would be interesting to analyze the formation of superradiant modes in this case, we will restrict our analysis to flat compact extra-dimensions, where a small but non-vanishing field mass is thus required for astrophysically relevant superradiant instabilities.

For the simplest case of black strings or branes at rest in the extra-dimensions, a bare field mass and an induced KK-mass are formally identical, although phenomenologically distinct, and the results of \cite{Cardoso:2004zz, Cardoso:2005vk} apply. However, as we show in this work, this is no longer the case when there is a non-vanishing KK-momentum along the compact directions, which modifies both the condition for superradiant scattering and the growth rate of the associated instability. As we discuss below, boosted black strings or branes may result from KK-number violating processes, so that our results suggest that superradiant instabilities may also be used to probe non-trivial compactification schemes.

Our analysis is done mostly in the context of Kerr black strings in five dimensions, extending the analysis in \cite{Dias:2006zv} for the case of massive fields, but it should be straightforward to extend our results to higher-dimensional boosted black branes. This work is organized as follows. In the next section we introduce the metric and the main properties of the boosted Kerr black string geometry. We derive the massive Klein-Gordon equation in this background in section III and separate it into its radial and angular components. Sections IV and V are devoted to the study of the radial equation using analytical and numerical methods, respectively, to determine the spectrum of quasi-bound states in this geometry. We discuss the potential phenomenological impact of our results in section VI, and summarize our main conclusions in section VII.

%%%%%%%%%%%%%%%%%%%%%%%%%%%%%%%%%%%%%%%%%%%%%%%%%%%%%%%%%%%%%%%%%%%%%%%%%%%%%%%%%%%%%%%%%%%%%%%%%%%%%%%%%%%%%%%%%%%%%%%%%%%%%%%%%%%%%%%%%%%%%%%%%%%%%%%%%%%%%%%%%%%%%%%%%%%%%%%%%%%%%%%%%%%%%%%%%%%%%%%%%%%%%%%%%%%%%%%%%%%%%%%%%%%%%%%%%%%%%%%%%%%%%%%%%%%%%%%%%%%%%%%%%%%%%%%%%%%%%%%%%%%%%%%%%%%%%%%%%%%%%%%%%%%%%%%%%%%%%%%%%%%%%%%%%%%%%%%%%%%%%%%%%%%%%%%%%%%%%%%%%%%%%%%%%%%%%%%%%%%%%%%%%%%%%%%%%%%%%%%%%%%%%%%%%%%%%%%%%%%%%%%%%%%%%%%%%%%%%%%%%%%%%%%%%%%%%%%%%%%%%%%%%%%%%%%%%%%%%%%%%%%%%%%%%%%%%%%%%%%%%%%%%%%%%%%%%%%%%%%%%%%%%%%%%%%%

\section{Boosted black strings}

A boosted Kerr black string in five dimensions is described by the line element, in Boyer-Lindquist coordinates \cite{Dias:2006zv}:
\begin{eqnarray} \label{string_metric}
ds^2&=&-\left(1-{2Mr\cosh^2\sigma\over \Sigma}\right)dt^2+{2Mr\sinh(2\sigma)\over\Sigma}dtdz+\left(1+{2Mr\sinh^2\sigma\over\Sigma}\right)dz^2+
{\Sigma\over\Delta}dr^2+\Sigma d\theta^2+\nonumber\\
&+&{(r^2+a^2)^2-\Delta a^2\sin^2\theta\over \Sigma}\sin^2\theta d\phi^2-{4Mr\cosh\sigma\over\Sigma}a\sin^2\theta dtd\phi-
{4Mr\sinh\sigma\over\Sigma}a\sin^2\theta dzd\phi~, 
\end{eqnarray}
where
\begin{equation} \label{metric_functions}
\Delta=r^2+a^2-2Mr~, \qquad \Sigma=r^2+a^2\cos^2\theta~,
\end{equation}
where we have set $G_5=\hbar=c=1$, which will be the case henceforth unless explicitly stated. The fifth dimension $z$ is assumed to be compact with size $L$ and $\sigma$ denotes the boost parameter along this direction, such that for $\sigma=0$ we obtain the rotating black string $ds^2=ds_{\mathrm{Kerr}}^2+dz^2$, characterized by a mass density $M$ and spin parameter $a\leq M$ to avoid naked singularities. 

This geometry has a curvature singularity at $r=0$ and inner (Cauchy) and outer horizons at $r_\pm=M\pm\sqrt{M^2-a^2}$, with an ergosurface at $r_e=M\cosh^2\sigma+\sqrt{M^2\cosh^4\sigma-a^2\cos^2\theta}$~. The angular and linear velocities of the black string at the horizon $r_+$ are given by:
\begin{eqnarray} \label{velocities}
\Omega={a\cosh\sigma\over r_+^2+a^2}~,\qquad  V_z=-\tanh\sigma. 
\end{eqnarray}

Note that the boost is not an observer-dependent quantity as the global Lorentz boost symmetry is broken for a compact direction, corresponding to the KK-charge of the black string \cite{Lee:2007bi}. We can write the total mass, angular momentum and linear momentum along the compact direction of the black string as:
\begin{eqnarray} \label{observables}
M_{BS}&=&ML\left({\cosh^2\sigma+1\over 2}\right)~,\nonumber\\
J_{BS}&=&aML\cosh\sigma~,\\
P_{BS}&=&{ML\over4}\sinh(2\sigma)\nonumber~. 
\end{eqnarray}
Although the linear momentum is quantized along the compact direction, given the large hierarchy expected between the size of the horizon for astrophysical black holes and the size of any compact extra-dimensions, we may take $\sigma$ as a continuous parameter.
Black strings and branes have also been shown to suffer from the Gregory-Laflamme instability, corresponding to long-wavelength longitudinal modes that grow without bound. For compact extra-dimensions, however, these modes are absent if the size of the extra-dimensions is small compared to the horizon radius, $L\lesssim r_+$, so that we expect astrophysical black holes to be stable against this type of perturbation \cite{Gregory:1993vy, Gregory:1994bj, Hovdebo:2006jy, Yoo:2011vu}.

%%%%%%%%%%%%%%%%%%%%%%%%%%%%%%%%%%%%%%%%%%%%%%%%%%%%%%%%%%%%%%%%%%%%%%%%%%%%%%%%%%%%%%%%%%%%%%%%%%%%%%%%%%%%%%%%%%%%%%%%%%%%%%%%%%%%%%%%%%%%%%%%%%%%%%%%%%%%%%%%%%%%%%%%%%%%%%%%%%%%%%%%%%%%%%%%%%%%%%%%%%%%%%%%%%%%%%%%%%%%%%%%%%%%%%%%%%%%%%%%%%%%%%%%%%%%%%%%%%%%%%%%%%%%%%%%%%%%%%%%%%%%%%%%%%%%%%%%%%%%%%%%%%%%%%%%%%%%%%%%%%%%%%%%%%%%%%%%%%%%%%%%%%%%%%%%%%%%%%%%%%%%%%%%%%%%%%%%%%%%%%%%%%%%%%%%%%%%%%%%%%%%%%%%%%%%%%%%%%%%%%%%%%%%%%%%%%%%%%%%%%%%%%%%%%%%%%%%%%%%%%%%%%%%%%%%%%%%%%%%%%%%%%%%%%%%%%%%%%%%%%%%%%%%%%%%%%%%%%%%%%%%%%%%%%%%

\section{Massive Klein-Gordon equation}

For a scalar field of mass $\mu$, the Klein-Gordon equation $(\nabla_\mu\nabla^{\mu}-\mu^2)\Phi=0$ in the background geometry given by Eq.~(\ref{string_metric}) can be separated into its angular and radial parts for perturbation modes of the form:
\begin{equation} \label{separation}
\Phi_{\Lambda}(t,r,\theta,\phi,z)=e^{i(-\omega t+m\phi-kz)}S_{lm}(\theta)R_{nlm}(r)~, 
\end{equation}
where $\Lambda=(\omega,k,l,m,n)$ describe the quantum numbers of the mode, yielding the equations:
\begin{equation} \label{angular_eq}
{1\over\sin\theta}\partial_\theta\left(\sin\theta\partial_\theta S\right)+\left[a^2(\omega^2-\mu^2-k^2)\cos^2\theta-{m^2\over\sin^2\theta}+\lambda\right]S=0~, 
\end{equation}
\begin{eqnarray} \label{radial_eq}
& &\Delta\partial_r\left(\Delta\partial_r R\right)-\Delta\left[(\mu^2+k^2)r^2+a^2\omega^2-2\omega mac_\sigma+\lambda\right]R+\nonumber\\
&+&\left[\left(\omega(r^2+a^2)-mac_\sigma\right)^2
+2Mr(r^2+a^2)c_\sigma^2(\omega-kt_\sigma)^2-2Mr(r^2+a^2)\omega^2-m^2a^2s_\sigma^2+4kmaMrs_\sigma\right]R=0~, \nonumber
\end{eqnarray}
where, for simplicity, we have dropped the labels $(n,l,m)$ and used the notation $c_\sigma=\cosh\sigma$, $s_\sigma=\sinh\sigma$ and $t_\sigma=\tanh\sigma$. The separation constant corresponds to the eigenvalue of the angular spheroidal harmonic equation (\ref{angular_eq}) and is given by:
\begin{equation} \label{angular_eigenvalue}
 \lambda=l(l+1)+\sum_{j=1}^{\infty}c_{jlm}(aq)^{2j}~,
\end{equation}
for $q=\sqrt{\mu^2+k^2-\omega^2}$, where the coefficients $c_{jlm}$ may be found in \cite{Abramowitz}. As we will show below, we are interested in quasi-bound state modes for which $q\ll1$, so that $\lambda=l(l+1)$ is generically a good approximation. 

As one can see from the radial equation, although each mode has an effective mass $\mu^2+k^2$ with contributions from both the bare mass and the KK-momentum, the latter has additional effects for a non-vanishing boost, so that we do not expect the bare and the KK-masses to affect the spectrum of perturbations in the same way. One should also note that, due to the translational symmetry of the line element Eq.~(\ref{string_metric}) along the $z$-coordinate, these equations hold for both a bulk or a localized scalar field, with $k=0$ in the latter case.

Before we compute the spectrum of quasi-bound states, it is useful to write the radial equation in terms of a re-scaled coordinate 
\begin{equation} \label{x_coordinate}
 x={r-r_+\over r_+}~,
\end{equation}
giving:
\begin{equation} \label{radial_eq_x}
x^2(x+\tau)^2\partial_x^2R+x(x+\tau)(2x+\tau)\partial_x R+V(x)R=0~, 
\end{equation}
where
\begin{eqnarray} \label{radial_potential}
V(x)&=&\left[x(x+2)\bar\omega+(2-\tau)(\bar\omega-m\bar\Omega)\right]^2+x(x+\tau)\bigg[(\tau-1)\bar\omega^2+2(2-\tau)\bar\omega m\bar\Omega-(\bar{k}^2+\bar\mu^2)(x+1)^2-\lambda\bigg]+\nonumber\\
&+&(2-\tau)(x+1)\left((x+1)^2+1-\tau\right)\left(c_\sigma^2(\bar\omega-t_\sigma\bar{k})^2-\bar\omega^2\right)-(2-\tau)^2m^2\bar\Omega^2t_\sigma^2 +2(2-\tau)^2(x+1)m\bar\Omega\bar{k}t_\sigma~.\nonumber\\
\end{eqnarray}
Here we have defined $\tau=(r_+-r_-)/r_+$, such that $\tau=0$ corresponds to an extremal and $\tau=1$ to a non-rotating black string. The barred dimensionless quantities are defined as $\bar\alpha=\alpha r_+$. In the next sections we will use analytical and numerical methods to determine the spectrum of quasi-bound states corresponding to this equation.

%%%%%%%%%%%%%%%%%%%%%%%%%%%%%%%%%%%%%%%%%%%%%%%%%%%%%%%%%%%%%%%%%%%%%%%%%%%%%%%%%%%%%%%%%%%%%%%%%%%%%%%%%%%%%%%%%%%%%%%%%%%%%%%%%%%%%%%%%%%%%%%%%%%%%%%%%%%%%%%%%%%%%%%%%%%%%%%%%%%%%%%%%%%%%%%%%%%%%%%%%%%%%%%%%%%%%%%%%%%%%%%%%%%%%%%%%%%%%%%%%%%%%%%%%%%%%%%%%%%%%%%%%%%%%%%%%%%%%%%%%%%%%%%%%%%%%%%%%%%%%%%%%%%%%%%%%%%%%%%%%%%%%%%%%%%%%%%%%%%%%%%%%%%%%%%%%%%%%%%%%%%%%%%%%%%%%%%%%%%%%%%%%%%%%%%%%%%%%%%%%%%%%%%%%%%%%%%%%%%%%%%%%%%%%%%%%%%%%%%%%%%%%%%%%%%%%%%%%%%%%%%%%%%%%%%%%%%%%%%%%%%%%%%%%%%%%%%%%%%%%%%%%%%%%%%%%%%%%%%%%%%%%%%%%%%%

\section{Analytic results}

Although the radial equation (\ref{radial_eq_x}) does not have an exact analytical solution, for small masses $\bar\mu\ll1$, $\bar{k}\ll1$ we may use Starobinsky's matching procedure \cite{Starobinsky} that has been employed in earlier analysis of superradiant instabilities (see e.g.~\cite{Cardoso:2004nk, Dias:2006zv, Rosa}). This method consists in dividing the exterior of the black hole into two overlapping regions where the radial equation can be solved analytically -  a near-horizon region, defined by $\bar\omega x\ll l$, and a far region, $x\gg1$ - and then match the solutions in their common domain of validity, $1\ll x\ll l/\bar\omega$. For quasi-bound states, we have $\omega\sim\sqrt{\mu^2+k^2}$, which justifies the validity of this procedure in the small mass limit.

Let us now derive the solutions of the radial equation in both regions. In the near-horizon region, we have:
\begin{eqnarray} \label{radial_potential_near}
V(x)\simeq \bar\varpi^2-\lambda x(x+\tau)~,
\end{eqnarray}
where 
\begin{eqnarray} \label{superradiant_omega_bar}
\bar\varpi=(2-\tau)c_\sigma\left(\bar\omega-{m\bar\Omega\over c_\sigma^2}-\bar{k}t_\sigma\right)~.
\end{eqnarray}
This generalizes the factor $\bar\varpi=(2-\tau)(\bar\omega-m\bar\Omega)$ for 4D Kerr black holes which determines the condition for superradiant scattering, $\bar\varpi<0$. The near-horizon region equation can then be reduced to the hypergeometric equation, and imposing `ingoing' boundary conditions at the horizon, $x=0$, we have:
\begin{equation} \label{near_solution}
R_{near}(x)=A\left(x\over x+\tau\right)^{-i\bar\varpi/\tau}{}_2F_1(l+1,-l,1-2i\bar\varpi,-x/\tau)~.
\end{equation}
Using the asymptotic properties of the hypergeometric function \cite{Abramowitz}, we may take the $x\gg\tau$ limit of this solution to obtain:
\begin{equation} \label{near_solution_large}
R_{near}(x)\simeq A\Gamma(1-2i\bar\varpi)\left[{\Gamma(2l+1)\over\Gamma(l+1)\Gamma(l+1-2i\bar\varpi)}\left({x\over \tau}\right)^{l}+{\Gamma(-2l-1)\over\Gamma(-l)\Gamma(-l-2i\bar\varpi)}\left({x\over \tau}\right)^{-l-1}\right]~.
\end{equation}

In the far region, $x\gg1$, the radial equation reduces to:
\begin{equation} \label{far_solution_eq}
x^2\partial_x^2R+2x\partial_xR+(-\bar{q}^2x^2+2\bar{q}\nu x-\lambda)R=0~,
\end{equation}
where
\begin{equation} \label{nu}
\nu={(2-\tau)\over2}{(\bar\omega c_\sigma-\bar{k}s_\sigma)^2-\bar{q}^2\over\bar{q}}~.
\end{equation}
The solution which is regular at infinity, corresponding to a bound state, is then given in terms of a confluent hypergeometric function:
\begin{equation} \label{far_solution}
R_{far}(x)=Bx^le^{-\bar{q}x}U(l+1-\nu,2l+2,2\bar{q}x)~.
\end{equation}
This reduces in the limit $\bar{q}x\ll1$ to \cite{Abramowitz}:
\begin{equation} \label{far_region_small}
R_{far}(x)\simeq B{\pi\over\sin((2l+2)\pi)}\left[{x^l\over\Gamma(-l-\nu)\Gamma(2l+2)}-{(2\bar{q})^{-(2l+1)}x^{-l-1}\over\Gamma(l+1-\nu)\Gamma(-2l)}\right]~.
\end{equation}
Hence, as expected, the solutions exhibit the same behaviour in their common domain of validity, and one can obtain the spectrum of bound state frequencies by matching the coefficients of $x^l$ and $x^{-l-1}$ in Eqs.~(\ref{near_solution_large}) and (\ref{far_region_small}), giving the condition:
\begin{equation} \label{matching_condition}
{\Gamma(-l-\nu)\Gamma(2l+2)\over\Gamma(l+1-\nu)\Gamma(-2l)}=-(2\bar{q}\tau)^{2l+1}{\Gamma(-2l-1)\over\Gamma(-l)}{\Gamma(l+1)\over\Gamma(2l+1)}{\Gamma(l+1-2i\bar\varpi)\over\Gamma(-l-2i\bar\varpi)}~.
\end{equation}
Notice that this condition exhibits several singular $\Gamma$-functions, but only due to the fact that we have neglected the higher-order corrections to the angular eigenvalue in Eq.~(\ref{angular_eigenvalue}), and in fact these poles cancel as we shall see below\footnote{Note that this cancellation does not occur for all choices of the near and far regions, as shown in \cite{Rosa} for extremal Kerr black holes, and that this may lead to unphysical enhancements of the growth rate.}. Taking into account that $\bar{q}\ll1$ for bound states in the small mass limit, to leading order the right-hand side of the matching condition vanishes, and we have the condition:
\begin{equation} \label{leading_condition}
\nu_0=l+1+n~,
\end{equation}
where $n$ is a non-negative integer, which corresponds to a pole in $\Gamma(l+1-\nu)$ that makes the left-hand side of the matching condition vanish to leading order. Using the definition of $\nu$ we may solve this for the bound state frequency, which we may write approximately as $\bar\omega=\bar\omega_0+\delta\bar\omega$, with:
\begin{equation} \label{leading_spectrum_1}
\bar\omega_0=\sqrt{\bar{\mu}^2+\bar{k}^2}~, \qquad \delta\bar\omega=-\left({2-\tau\over2}\right)^2{(\bar\omega_0 c_\sigma-\bar{k}s_\sigma)^4\over2\bar\omega_0(l+1+n)^2}~.
\end{equation}
We are particularly interested in modes with $k=0$, as discussed above, in which case this yields the Hydrogen-like spectrum typical of massive fields in a gravitational potential:
\begin{equation} \label{leading_spectrum_2}
\omega\simeq \mu\left(1-{(\mu M)^2c_\sigma^4\over2(l+1+n)^2}\right)~,
\end{equation}
where we have restored the original (unbarred) variables. We see that the leading effect of the boost along the compact direction is to make the states more tightly bound to the black hole, as compared to the unboosted case. 

To obtain the imaginary part of the bound state frequency, which determines whether the modes are superradiant and hence unstable or not, we need to compute the leading corrections from the non-zero value of the right-hand side of Eq.~(\ref{matching_condition}). To do this, we proceed as in \cite{Furuhashi:2004jk} and expand the left-hand side of the matching condition with $\nu=\nu_0+\delta\nu$ and evaluate the right-hand side with the leading order result $\omega_0$. We use the following results for cancelling the poles in the $\Gamma$-functions:
\begin{eqnarray} \label{Gamma_functions}
\lim_{z\rightarrow-n}{\psi(z)\over\Gamma(z)}=(-1)^{n+1}n!~,\qquad {\Gamma(-2l-1)\over \Gamma(-l)}={(-1)^{l+1}\over2}{l!\over (2l+1)!}~,\qquad
{\Gamma(l+1+x)\over\Gamma(-l-x)}=(-1)^lx\prod_{j=1}^l(j^2-x^2)~,
\end{eqnarray}
where $\psi(z)=\Gamma'(z)/\Gamma(z)$. Note that in deriving the second result one needs to take into account that the angular eigenvalue receives higher-order corrections, according to Eq.~(\ref{angular_eigenvalue}), so that one should replace $l\rightarrow l+\epsilon$ and take the limit $\epsilon\rightarrow0$, which yields the factor $1/2$ in this expression. After some algebra, we obtain:
\begin{equation} \label{imaginary_spectrum}
\mathrm{Im}(\omega M)=-{1\over2}C_{ln}(\varpi M)(\mu M)^{4l+5}c_\sigma^{4l+6}\left({r_+-r_-\over r_++r_-}\right)^{2l+1} ~,
\end{equation}
with
\begin{equation} \label{imaginary_coefficient}
C_{ln}={4^{2l+2}\over (l+1+n)^{2l+4}}{(2l+1+n)!\over n!}\left({l!\over(2l+1)!(2l)!}\right)^2\prod_{j=1}^l(j^2+16(\varpi M)^2)~.
\end{equation}
Also, for $k=0$, we have the unbarred quantity
\begin{equation} \label{superradiant_omega}
\varpi=c_\sigma\left(\omega-{m\Omega\over c_\sigma^2}\right)\simeq c_\sigma\left(\mu-{m\Omega\over c_\sigma^2}\right)~.
\end{equation}
This implies that superradiant bound states satisfy $\mu<m\Omega/c_\sigma^2$ in the small mass limit, so that the boost reduces the range of unstable modes. On the other hand, the boost increases the growth rate of the instability, as can be seen from the strong dependence of $\mathrm{Im}(\omega M)$ on $\cosh\sigma$. We have, for example, for $l=m=1$ and $n=0$, for slowly-rotating black strings with $a\ll M$ and $\mu\ll \Omega$:
\begin{equation} \label{superradiant_dipole}
\mathrm{Im}(\omega M)\simeq {1\over48}\left({a\over M}c_\sigma\right)(\mu M c_\sigma)^9~,
\end{equation}
which reproduces the result obtained for 4D Kerr black holes in the limit $\sigma=0$. Note that this differs from the results in \cite{Detweiler:1980uk, Furuhashi:2004jk} by a factor $1/2$, which is associated with taking into account the corrections to the angular eigenvalue in Eq.~(\ref{Gamma_functions}), in agreement with the recent analysis in \cite{Pani:2012fu}. This procedure does not allow one to explore the near-extremal limit, where $\tau\rightarrow0$, and where we expect the growth rate of the superradiant instability to be stronger, as is the case of 4D Kerr black holes \cite{Cardoso:2005vk, Dolan:2007mj, Rosa, Arvanitaki}. Also, it is clear from Eq.~(\ref{leading_spectrum_2}) that the approximations used to derive the superradiant spectrum break down for  large boosts. Numerical methods are thus required in order to determine the spectrum of bound states for general spin and boost parameters, which we will consider in the next section.

%%%%%%%%%%%%%%%%%%%%%%%%%%%%%%%%%%%%%%%%%%%%%%%%%%%%%%%%%%%%%%%%%%%%%%%%%%%%%%%%%%%%%%%%%%%%%%%%%%%%%%%%%%%%%%%%%%%%%%%%%%%%%%%%%%%%%%%%%%%%%%%%%%%%%%%%%%%%%%%%%%%%%%%%%%%%%%%%%%%%%%%%%%%%%%%%%%%%%%%%%%%%%%%%%%%%%%%%%%%%%%%%%%%%%%%%%%%%%%%%%%%%%%%%%%%%%%%%%%%%%%%%%%%%%%%%%%%%%%%%%%%%%%%%%%%%%%%%%%%%%%%%%%%%%%%%%%%%%%%%%%%%%%%%%%%%%%%%%%%%%%%%%%%%%%%%%%%%%%%%%%%%%%%%%%%%%%%%%%%%%%%%%%%%%%%%%%%%%%%%%%%%%%%%%%%%%%%%%%%%%%%%%%%%%%%%%%%%%%%%%%%%%%%%%%%%%%%%%%%%%%%%%%%%%%%%%%%%%%%%%%%%%%%%%%%%%%%%%%%%%%%%%%%%%%%%%%%%%%%%%%%%%%%%%%%%

\section{Numerical results}

The spectrum of quasi-bound states can be computed using a simple numerical method, which takes advantage of the fact that the solutions decay exponentially far away from the black hole horizon, and was employed in this context in \cite{Rosa:2011my}. Near the horizon of the black hole, $x\ll1$ ,the mode functions can be expanded in a Taylor series of the form:
\begin{equation} \label{near_horizon_expansion}
R(x)=x^{-i\bar\varpi/\tau}\sum_{n=0}^{\infty}a_nx^n~. 
\end{equation}
The $a_n$ coefficients may be easily computed by replacing this {\it ansatz} into Eq.~(\ref{radial_eq_x}) and using a symbolic algebraic package such as Mathematica to solve it order by order. The overall normalization is irrelevant in determining the spectrum, so we can set $a_0=1$. 

We may then use the obtained series to set the boundary condition at $x=\epsilon\ll1$ and numerically integrate Eq.~(\ref{radial_eq_x}) up to an arbitrarily large distance from the horizon, $x=d$, as a function of the frequency $\omega$ for each $(l,m)$ mode. Bound states will then correspond to the modes that minimize $R(d)$ in the complex $\omega$-plane, and this can also be determined by using the basic numerical minimization tools available in Mathematica or other similar packages. An arbitrary degree of precision may then be achieved by computing an increasing number of terms in the near-horizon expansion and taking large values of the distance $d$, but this is of course limited by the numerical precision of the code and computational time constraints. For the minimization procedure we use the leading order result in Eq.~(\ref{leading_spectrum_1}) as a starting point.

In Figure 1 we show the results obtained for imaginary part of the fastest growing mode, $l=m=1$, $n=0$, as a function of the mass coupling $\mu M$ and for different values of the spin $a/M$ and boost parameter $\cosh\sigma$ of the rotating black string\footnote{The value $\cosh\sigma=\sqrt2$ corresponds to the large radius limit of a balanced doubly rotating thin black ring analyzed in \cite{Dias:2006zv}.}.

\begin{figure}[htbp]
\centering\includegraphics{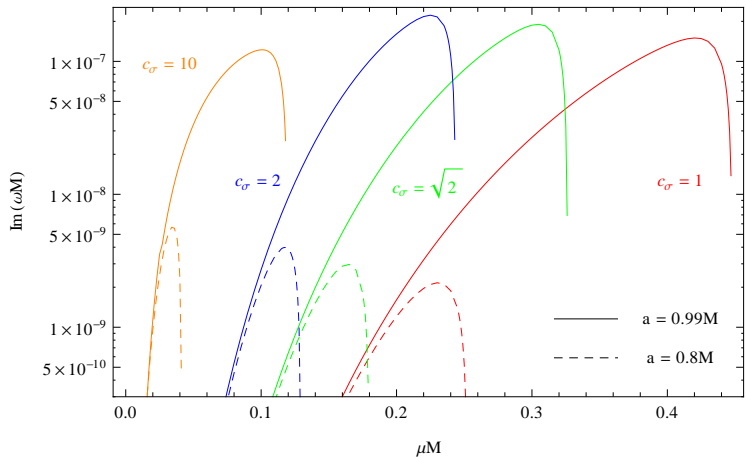} 
\caption{Imaginary part of the bound state frequency as a function of the dimensionless mass coupling $\mu M$ for $a=0.99M$ and $a=0.8M$ and different values of the boost parameter, $c_\sigma\equiv\cosh\sigma$.}
\end{figure}

We hence see that, as suggested by our approximate analytical results, if a black string carries a non-vanishing KK-momentum, the spectrum of superradiant bound states changes significantly, first by reducing the maximum field mass for which the instability is present but also by modifying the growth rate of the unstable modes. For $\sigma=0$, we obtain a maximum growth rate for $a/M=0.99$ of $\mathrm{Im}(\omega M)\simeq 1.5\times 10^{-7}$, in agreement with \cite{Cardoso:2005vk, Dolan:2007mj, Rosa}. The results in this figure suggest that the maximum growth rate for near-extremal geometries does not have a monotonic behaviour, increasing for small values of $c_\sigma$ but decreasing at large KK-momentum. To better understand this behaviour, we have determined numerically the maximum growth rate as a function of the boost parameter, which is illustrated in Figure 2.

\begin{figure}[htbp]
\centering\includegraphics[scale=1]{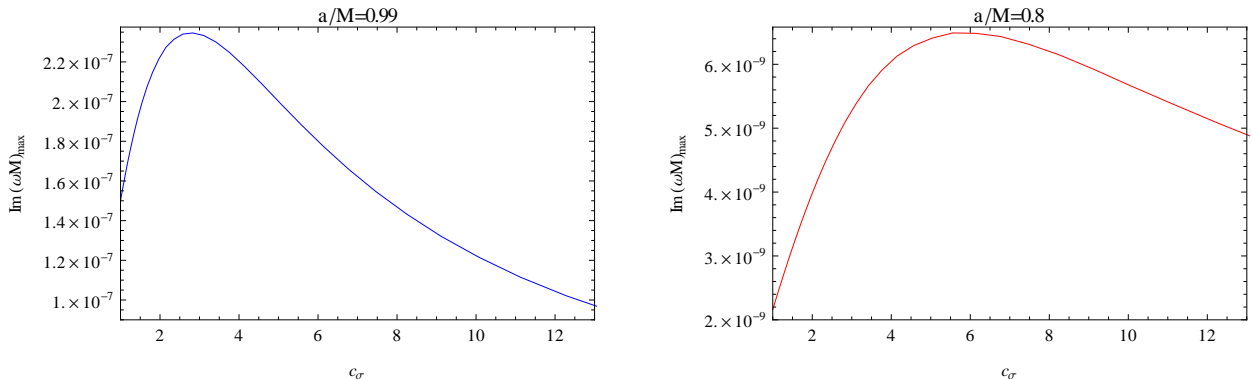} 
\caption{Maximum value of the imaginary part of the bound state frequency as a function of the boost parameter, $c_\sigma\equiv\cosh\sigma$, for $a=0.99M$ (left) and $a=0.8M$ (right).}
\end{figure}

These results confirm our expectations from the analytical study in the previous section, namely that for small values of the boost parameter the maximum growth rate of the instability increases with $c_\sigma$, despite the decrease in the upper limit of the superradiant region. However, there is a maximum enhancement for a given value of $a/M$ and a critical boost parameter above which the superradiant instability becomes suppressed. This behaviour can be understood by analyzing the corresponding values of the real part of the bound state frequency, given in Figure 3.

\begin{figure}[htbp]
\centering\includegraphics[scale=1]{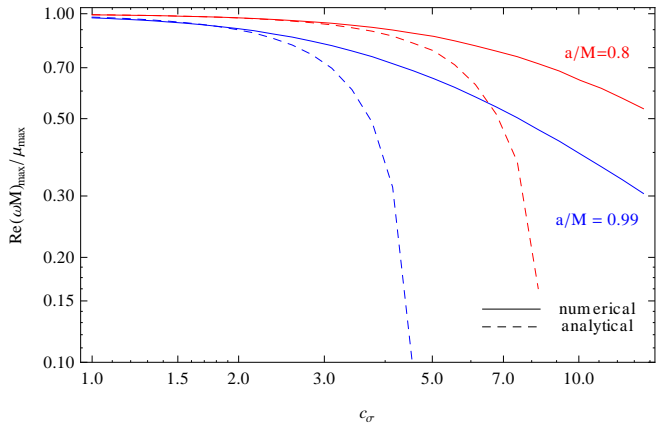} 
\caption{Results obtained for the normalized real part of the bound state frequency at the field mass value corresponding to the maximum superradiant growth rate, as a function of the boost parameter, $c_\sigma\equiv\cosh\sigma$. The solid lines correspond to the numerically obtained solution, while the dashed lines yield the Hydrogen-like spectrum in Eq.~(\ref{leading_spectrum_2}).}
\end{figure}

We see from the results in Figure 3 that the critical value of the boost parameter above which the superradiant growth rate begins to decrease coincides approximately with the value at which the numerical results start to deviate significantly from a Hydrogen-like spectrum and, in particular, for large $c_\sigma$ the scalar states are not as tightly bound to the black string as previously estimated. This implies that the value of  $q=\sqrt{\mu^2-\omega^2}$ grows more slowly with the boost parameter than the analytical procedure suggested, which in practice cannot compensate the decrease in the upper limit of the superradiant spectrum. In the light of the tunneling picture devised in \cite{Arvanitaki-JMR}, the parameter $q^{-1}$ determines the height of the potential barrier separating the potential well created by the field mass and the ergoregion of the black string where superradiant amplification occurs. A smaller value of $q$ thus implies a smaller overlap of the bound state wavefunction with the ergoregion and hence a smaller amplification rate.

For practical purposes the maximum growth rate of the superradiant instability is well-described by an expression of the form:
\begin{equation} \label{maximum_growth}
\mathrm{Im}(\omega M)_{\mathrm{max}}\simeq {\alpha c_\sigma\over 1+\beta c_\sigma^2}~,
\end{equation}
where $\alpha$ and $\beta$ depend on the spin parameter $a/M$. For example, in the near-extremal case $a=0.99M$ this yields $\alpha=1.7\times 10^{-7}$ and $\beta=0.13$, with both parameters decreasing with the black string spin. This means that the superradiant growth rate can be up to $1.56$ times larger than the unboosted case for near-extremal geometries, with a stronger enhancement for smaller spin.

Finally, it is useful to compare our numerical results for the superradiant growth rate with the analytical prediction obtained in the previous section. We illustrate this in Figure 4 for two different values of the boost and spin parameters. 

\begin{figure}[htbp]
\centering\includegraphics[scale=0.82]{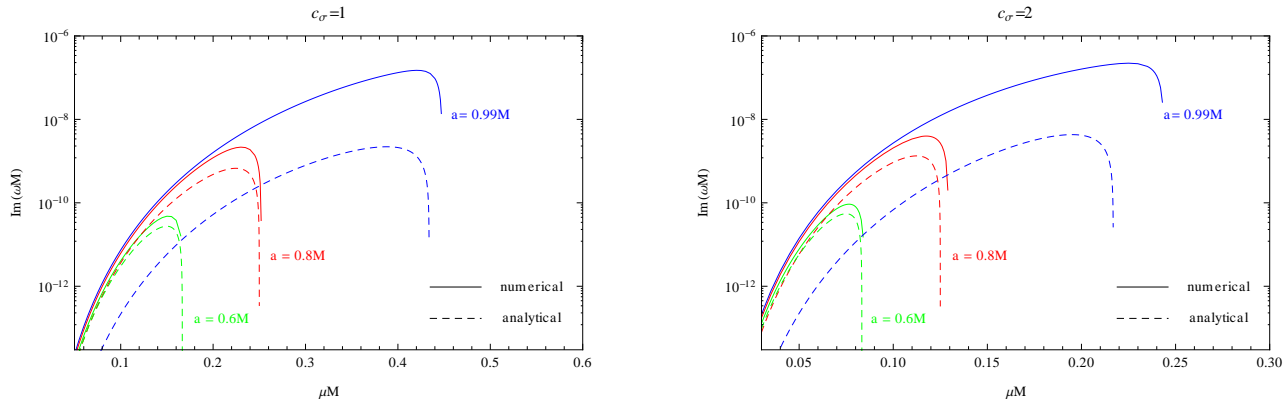} 
\caption{Imaginary part of the bound state frequency as a function of the dimensionless mass coupling $\mu M$ for different values of $a/M$, with $c_\sigma=1$ (left) and $c_\sigma=2$ (right). The solid lines correspond to the numerical results, while the dashed lines give the analytical prediction from the matching procedure in the small mass and boost limit.}
\end{figure}
  
We conclude that the analytical expression underestimates the growth rate of the superradiant instability for large masses and spin $a/M$, the deviations growing as one increases the spin of the black string. The endpoint of the instability is also underestimated by the analytical prediction, in particular for larger boost $c_\sigma$. It is interesting to note that a much better agreement is obtained if one removes the $l$-dependence of the factor $(r_+-r_-/r_++r_-)^{2l+1}$ appearing in Eq.~(\ref{imaginary_spectrum}), i.e. setting $l=0$. This is illustrated in Figure 5, and although it gives an empirical formula that does not follow from any rigorous computation, it is useful for practical purposes. 

\begin{figure}[htbp]
\centering\includegraphics[scale=0.82]{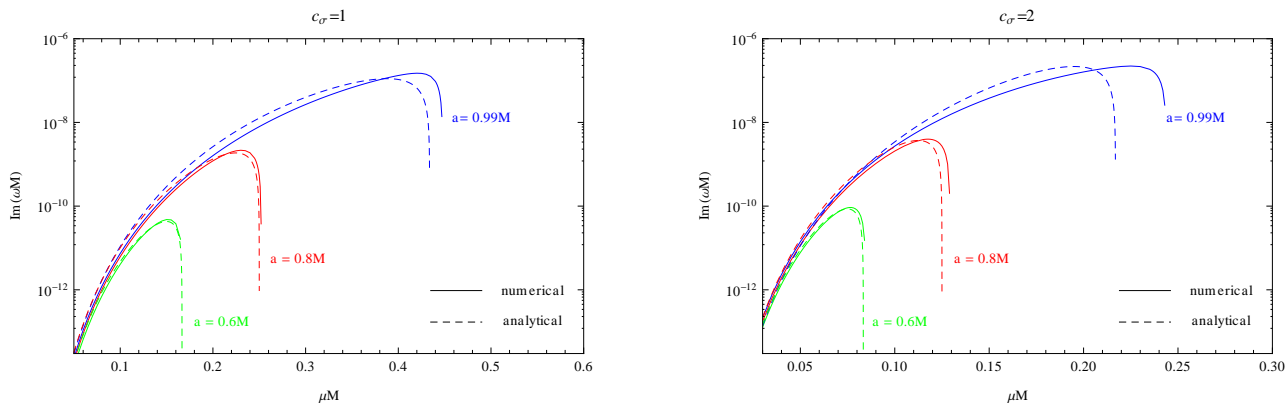} 
\caption{Imaginary part of the bound state frequency as a function of the dimensionless mass coupling $\mu M$ for different values of $a/M$, with $c_\sigma=1$ (left) and $c_\sigma=2$ (right). The solid lines correspond to the numerical results, while the dashed lines give the analytical prediction from the matching procedure in the small mass and boost limit, setting $l=0$ in the factor $(r_+-r_-/r_++r_-)^{2l+1}$ appearing in Eq.~(\ref{imaginary_spectrum}).}
%in Eq.~(\ref{imaginary_spectrum}).}
\end{figure}

%%%%%%%%%%%%%%%%%%%%%%%%%%%%%%%%%%%%%%%%%%%%%%%%%%%%%%%%%%%%%%%%%%%%%%%%%%%%%%%%%%%%%%%%%%%%%%%%%%%%%%%%%%%%%%%%%%%%%%%%%%%%%%%%%%%%%%%%%%%%%%%%%%%%%%%%%%%%%%%%%%%%%%%%%%%%%%%%%%%%%%%%%%%%%%%%%%%%%%%%%%%%%%%%%%%%%%%%%%%%%%%%%%%%%%%%%%%%%%%%%%%%%%%%%%%%%%%%%%%%%%%%%%%%%%%%%%%%%%%%%%%%%%%%%%%%%%%%%%%%%%%%%%%%%%%%%%%%%%%%%%%%%%%%%%%%%%%%%%%%%%%%%%%%%%%%%%%%%%%%%%%%%%%%%%%%%%%%%%%%%%%%%%%%%%%%%%%%%%%%%%%%%%%%%%%%%%%%%%%%%%%%%%%%%%%%%%%%%%%%%%%%%%%%%%%%%%%%%%%%%%%%%%%%%%%%%%%%%%%%%%%%%%%%%%%%%%%%%%%%%%%%%%%%%%%%%%%%%%%%%%%%%%%%%%%%

\section{Dimensional reduction}

In the previous sections we have seen that the non-vanishing boost modifies the spectrum of scalar superradiant bound states with respect to a black string geometry at rest. To better understand the origin of these changes, it is convenient to study the effective four-dimensional geometry corresponding to a boosted black string, which was analyzed for arbitrary dimensions in \cite{Kunz:2006jd}. The five-dimensional metric in Eq.~(\ref{string_metric}) can then be decomposed as:
\begin{equation} \label{metric_decomposition}
ds^2=e^{2\alpha\varphi}ds_4^2+e^{-4\alpha\varphi}(dz+A_\mu dx^\mu)^2~, 
\end{equation}
where $\varphi$ is the dilaton and $A_\mu$ the `gravi-photon' field, with $\alpha^{-1}=2\sqrt{3}$ yielding the canonical Einstein-Maxwell-dilaton action in four dimensions. Comparing Eqs.~(\ref{string_metric}) and (\ref{metric_decomposition}), one obtains:
\begin{eqnarray} \label{decomposed_quantities}
ds_4^2&=&e^{-2\alpha\varphi}\left[-dt^2+{\rho^2\over\Delta}dr^2+\rho^2d\theta^2+(r^2+a^2)\sin^2\theta d\phi^2+e^{4\alpha\varphi}{2Mr\over \rho^2}\left(c_\sigma dt-a\sin^2\theta d\phi\right)^2\right]~,\nonumber\\
A_\mu dx^\mu&=&{2Mr\over \rho^2}s_\sigma e^{4\alpha\varphi}\left(c_\sigma dt-a\sin^2\theta d\phi\right)~,\\
e^{-4\alpha\varphi}&=&1+{2Mr\over\rho^2}s_\sigma^2~.\nonumber
\end{eqnarray}
It is easy to see that for $\sigma=0$ one obtains the standard four-dimensional Kerr metric with trivial vector and dilaton fields. The solution for a non-vanishing boost is quite similar to the Kerr-Newman solution, but it yields important differences. In particular, this solution is not only characterized by a non-vanishing `gravi-electric' charge corresponding to the KK-momentum along the compact direction, $Q_{BS}=P_{BS}$, but also by a `gravi-magnetic' moment and dilatonic charge given, respectively, by:
\begin{eqnarray} \label{charges}
\mathcal{M}_{BS}=as_\sigma {ML\over 2}~,\nonumber\\
\Phi_{BS}={s_\sigma^2\over 8\alpha}ML~.
\end{eqnarray}

The spectrum of superradiant bound states for a rotating black hole with a non-vanishing electric charge has been studied by Furuhashi and Nambu in \cite{Furuhashi:2004jk}, where it was shown that the superradiant condition is shifted with respect to the uncharged case, yielding:
\begin{equation} \label{superradiant_condition_Kerr_Newman}
\omega<m\Omega+q\tilde{Q}~, 
\end{equation}
where $\tilde{Q}=Q/2M$ is the normalized charge of the black hole and $q$ the charge of the scalar field. Comparing with Eq.~(\ref{superradiant_omega_bar}), this yields the identification $q=4k/c_\sigma^2$, so that as expected the KK-number corresponds to the effective `gravi-electric' charge of a given mode. However, we have seen that the superradiant condition for a boosted black string is modified even for zero-modes, i.e. uncharged fields. This is of course associated with the additional charges carried by the compactified boosted black string, in particular the non-trivial dilaton profile, which affects all modes irrespective of their KK-number. This also results in the observed change of the instability growth rate, which is thus distinct from the Kerr-Newman case.

One should note, however, that in realistic scenarios the extra-dimensional moduli, in particular the dilaton field, must be stabilized in order to avoid significant violations of the weak equivalence principle \cite{Taylor:1988nw}, deviations from general relativity \cite{Scherk:1974ca} or cosmological variation of fundamental constants \cite{Witten:1984dg}. This generically leads to a heavy dilaton e.g. with a mass of the order of the supersymmetry breaking scale in the context of string theory and supergravity scenarios. Static black hole solutions in the presence of a massive dilaton have been discussed in \cite{Gregory:1992kr, Horne:1992bi}, where it was shown that the standard Reissner-Nordstr\"om solution is obtained when the Compton wavelength of the dilaton is much smaller than the horizon radius, whereas for small black holes the massless dilaton solution corresponding to the non-rotating limit of Eq.~(\ref{decomposed_quantities}) is recovered. As far as we are aware, no study of the rotating case in the presence of a massive dilaton has been performed so far and it is unclear whether the standard Kerr-Newman solution is obtained for large black holes. It would thus be interesting to investigate how the development of superradiant instabilities differs in the massive and massless dilaton cases, the latter corresponding to the compactified boosted black string analyzed in this work and first obtained in \cite{Horne:1992zy}. However, it has been shown by Damour and Polyakov \cite{Damour:1994zq} that string-loop corrections to the dilaton-matter couplings may lead to the dynamical stabilization of a massless dilaton through its cosmological evolution in a way consistent with experimental data, so that Eq.~(\ref{decomposed_quantities}) could correspond to a realistic solution in this scenario. While these are important open questions that need to be addressed, we will focus the remainder of our discussion on such a stable massless dilaton scenario.

%%%%%%%%%%%%%%%%%%%%%%%%%%%%%%%%%%%%%%%%%%%%%%%%%%%%%%%%%%%%%%%%%%%%%%%%%%%%%%%%%%%%%%%%%%%%%%%%%%%%%%%%%%%%%%%%%%%%%%%%%%%%%%%%%%%%%%%%%%%%%%%%%%%%%%%%%%%%%%%%%%%%%%%%%%%%%%%%%%%%%%%%%%%%%%%%%%%%%%%%%%%%%%%%%%%%%%%%%%%%%%%%%%%%%%%%%%%%%%%%%%%%%%%%%%%%%%%%%%%%%%%%%%%%%%%%%%%%%%%%%%%%%%%%%%%%%%%%%%%%%%%%%%%%%%%%%%%%%%%%%%%%%%%%%%%%%%%%%%%%%%%%%%%%%%%%%%%%%%%%%%%%%%%%%%%%%%%%%%%%%%%%%%%%%%%%%%%%%%%%%%%%%%%%%%%%%%%%%%%%%%%%%%%%%%%%%%%%%%%%%%%%%%%%%%%%%%%%%%%%%%%%%%%%%%%%%%%%%%%%%%%%%%%%%%%%%%%%%%%%%%%%%%%%%%%%%%%%%%%%%%%%%%%%%%%%

\section{Observational prospects}

As argued earlier, one expects black string/brane-like configurations to describe large astrophysical black holes in a universe with small and compact extra-dimensions. Given the results derived in the previous section, it is interesting to discuss the possibility that such black strings may carry additional charges, in particular KK-momentum. 

In realistic astrophysical environments, it is widely believed that, due to the strength of electromagnetic interactions, a black hole should maintain the electric charge neutrality of the parent star, or that a highly charged black hole would be quickly depleted of its charge by interactions with matter in its vicinity (see e.g. \cite{Carroll}). Since this is related to charge conservation in electromagnetic processes, one should ask whether this is also the case for KK-charges.

Although Lorentz invariance in our macroscopic (3+1)-dimensional world has been tested to a high degree of accuracy, momentum conservation in hidden dimensions is not necessarily a desirable feature of realistic scenarios. Firstly, the compactification itself breaks the translational symmetry of the extra-dimensions, inducing KK-momentum quantization. KK-number is nevertheless a conserved quantity if the extra-dimensions have an otherwise trivial structure. However, in string theory compactifications and related scenarios, one generically expects topological (and possibly dynamical) objects such as branes and orbifold/orientifold planes to further break the translational symmetry in the extra-dimensions. 

A particularly interesting scenario for our discussion is the case of Universal Extra-Dimensions (UED), where all Standard Model fields are allowed to propagate in the bulk of one or more extra-dimensions \cite{Appelquist:2000nn}. In order to obtain chiral zero-modes for quark and lepton fields, one generically considers orbifold compactifications, the simplest example being the five-dimensional $S^1/\mathbb{Z}_2$ orbifold, with fields being either even or odd under the orbifold symmetry transformation. These scenarios exhibit several attractive features from the phenomenological point of view, including a motivation for three fermion generations from anomaly cancellation \cite{Dobrescu:2001ae}, preventing rapid proton decay from non-renormalizable operators \cite{Appelquist:2001mj, ArkaniHamed:1999dc} and a natural explanation for the observed fermion mass hierarchy and mixings \cite{Mirabelli:1999ks, Dvali:2000ha, Kaplan:2000av, Kaplan:2001ga, Appelquist:2002ft}. Moreover, one of the most interesting motivations to consider UED scenarios lies in the existence of a natural dark matter candidate \cite{Servant:2002aq, Cheng:2002ej}. This relies on the assumption that the orbifold compactification does not completely break the translational symmetry in the extra-dimensions, in particular that potential KK-number-violating interactions at the orbifold fixed points (or branes) preserve a discrete KK-parity that ensures the stability of the lightest Kaluza-Klein particle (LKP), in a similar way to how R-parity conservation in supersymmetric models leads to the stability of the lightest supersymmetric particle (LSP). Accounting for radiative mass corrections \cite{Cheng:2002iz}, the most promising dark matter candidates in this context are the first KK-excitations of the photon or the neutrino, as well as possibly the lightest KK-graviton.

The existence of stable particles with a non-trivial KK-charge and accounting for a significant fraction of the dark matter in the universe is particularly interesting in the context of our earlier discussion, since astrophysical black holes would then naturally absorb large amounts of KK-momentum in a galactic and possibly extra-galactic environment. The orbifold symmetry implies, however, that the mass eigenstates are linear combinations of opposite KK-momentum states which are either even or odd under the orbifold transformation, so that even in the simplest case one expects the net KK-charge of an astrophysical black hole to vanish.

More interesting scenarios may arise, however, if one allows for additional KK-number-violating interactions that for example lead to the generation of a KK-charge asymmetry in the early universe, much in the same way that B- or L-violating interactions lead to the observed baryon asymmetry. Although to our knowledge no such models have been considered so far in the context of UED, asymmetric dark matter scenarios have attracted a significant interest in the recent literature, in particular in addressing the puzzling similarity between the observed dark and baryonic matter abundances \cite{Kaplan:2009ag}. As mentioned above, KK-number violating interactions may arise at the orbifold fixed points and possibly from interactions with fields which are localized in the extra-dimensions, as well as in scenarios where the Standard Model fields live a `thick brane' embedded in a higher-dimensional space, such that the thick brane absorbs the unbalanced KK-momentum \cite{Appelquist:2000nn}. If these interactions are sufficiently suppressed, they may be relevant only at the large temperatures occurring in the early universe and, despite explicitly breaking the residual KK-parity, ensure a cosmologically long-lived LKP that could account for the dark matter content of the universe. KK-number conservation would, in this case, also be preserved at the scales relevant for astrophysical black holes and, in particular, the development of superradiant instabilities. We would then expect such a cosmological KK-charge asymmetry of the universe to potentially induce a non-zero boost for astrophysical black strings/branes.

Although it is beyond the scope of this paper to build a concrete model implementing the generation of a KK-charge asymmetry, the above considerations motivate us to further explore the potential observational consequences of boosted black string bombs. While in the previous sections we have shown that KK-momentum changes the superradiant condition for a given value of $a/M$, one must take into account that neither this quantity nor the mass density $M$ are observable parameters, as the total mass and spin of the black string depend on the boost parameter, according to Eq.~(\ref{observables}). In order to determine whether KK-momentum may have any observable effect, one should first express the spectrum in terms of the physical quantities, $M_{BS}$ and $J_{BS}$.

Firstly, one should include the missing factors $G_5$, $\hbar$ and $c$. In particular, we have the dimensionless quantities:
\begin{eqnarray} \label{physical_quantities}
\mu M&\rightarrow& {\mu G_5M\over\hbar c}={2\over c_\sigma^2+1}{G_4M_{BS}\mu\over \hbar c}~,\nonumber\\
{a\over M}&\rightarrow &{ac\over G_5 M}={(c_\sigma^2+1)^2\over 4c_\sigma}{J_{BS}c\over G_4M_{BS}^2}~,\nonumber\\
{\omega M}&\rightarrow &{\omega G_5M\over c^3}={2\over c_\sigma^2+1}{G_4M_{BS}\omega\over c^3}~,
\end{eqnarray}
where we have used that $G_5=G_4L$ for the relation between the five-dimensional and four-dimensional Newton constants. Hence, as none of these quantities depends on the size of the extra-dimensions, we conclude that in the absence of KK-momentum the spectrum of superradiant bound states is indistinguishable from the four-dimensional case. The presence of a boost, on the other hand, has a potentially observable effect and, in particular, using $a\leq M$ ($ac\leq G_5M$) we derive the upper bound:
\begin{eqnarray} \label{ang_mom_bound}
{J_{BS}c\over G_4M_{BS}^2}\leq{4c_\sigma\over (c_\sigma^2+1)^2}~.
\end{eqnarray}
This bound is always smaller than unity and is strictly decreasing with $\sigma$, so that for example a highly boosted near-extremal black string could be mistaken for a slowly-rotating black hole.

One particular observational signature of the formation of superradiant bound states around black holes is the presence of gaps in the Regge plot, $(M_{BH},J_{BH}c/G_4M_{BH}^2)$. This is due to the fact that superradiant bound states extract angular momentum from the black hole, which competes with the spin-up process resulting from accretion of nearby matter. In \cite{Arvanitaki}, the authors estimate a minimum of $10^2$ e-folds for superradiant bound states to significantly reduce the angular momentum of a black hole, so that if the time-scale of the superradiant instability, $\tau_{sr}=\mathrm{Im}(\omega)^{-1}$, is at least one hundred times smaller than the Eddington accretion time, $\tau_E=\sigma_T/4\pi G m_P\sim 4\times 10^8 $ years, a black hole will be efficiently spun down. This condition thus defines an upper bound for the spin of a black hole, and we can easily extend this for the case of boosted Kerr black strings. 

We consider as an example the Regge plot corresponding to a QCD axion of mass $\mu=3\times 10^{-11}$ eV \cite{Arvanitaki} for different values of the boost parameter in Figure 6. Note that, for clarity of the plot, we have included only the fastest growing mode $l=m=1$, $n=0$, whereas the contributions of other superradiant modes should also be included. In constructing this plot, we have used the numerical solution of the massive Klein-Gordon equation for black string masses above the solar mass while, due to numerical precision, for lighter black holes we have used the analytical prediction in Eq.~(\ref{imaginary_spectrum}) setting $l=0$ in the factor $(r_+-r_-/r_++r_-)^{2l+1}$, which as shown above gives a better approximation to the numerical results.

\begin{figure}[htbp]
\centering\includegraphics[scale=1.15]{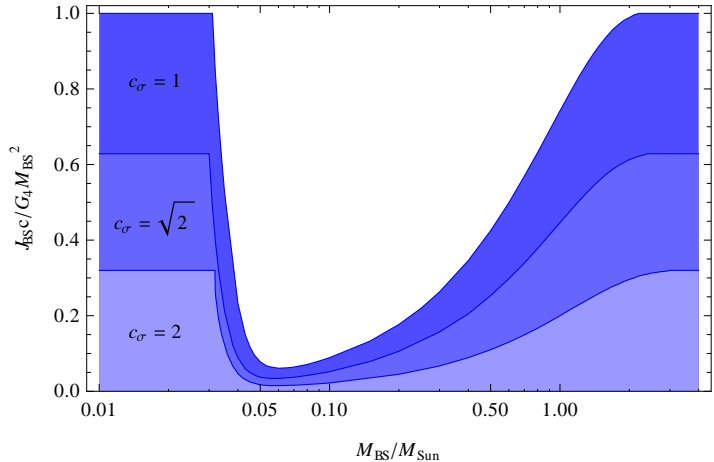} 
\caption{The black string Regge plot, showing the total mass $M_{BS}$ in units of the solar mass and the spin-mass ratio $J_{BS}c/G_4M_{BS}^2$, for different values of the boost parameter $c_\sigma\equiv\cosh\sigma$ and an axion with mass  $\mu=3\times 10^{-11}$ eV in the $l=m=1$, $n=0$ state. The colored regions correspond to the allowed spin-mass ratios in each case.}
\end{figure}

As one can see in this figure, the presence of KK-momentum leads to deeper and wider gaps in the black string Regge plot. This is mainly a consequence of the modified superradiant condition and of the upper bound on the physical spin-mass ratio of the black hole, being however less sensitive to the value of the maximum growth rate. The use of the analytical prediction in the small mass region implies that the gap is not accurately described for large spins, but this nevertheless describes well the main qualitative features of the Regge plot in this region.

If astrophysical black holes are characterized by a universal boost parameter, then we could expect to infer both the axion mass and the boost parameter from the shape of the Regge curve. In fact, a universal boost parameter could be determined independently of the existence of light scalar particles, if a consistent upper bound on the spin-mass ratio would be observed, although other astrophysical effects could {\it a priori} be responsible for such an effect. A universal boost parameter is, however, already significantly constrained by current observations of black hole candidates, such as GRS 1915+15, with $Jc/G_4M^2=0.98-1$ \cite{McClintock:2009dn}, which would imply $c_\sigma<1.01$ neglecting other astrophysical effects, corresponding to $P_{BS}/M_{BS}<0.07$. 

The case of non-universal boosts is probably more generic, and there are examples of astrophysical candidates with a small bound on the spin-mass ratio, such as LMC X-3, with $Jc/G_4M^2<0.26$ \cite{McClintock:2009dn}, which would allow for $c_\sigma\gtrsim 2$. It would be more difficult to determine the presence of a boost in this case, since the largest upper bound on the spin-mass ratio corresponds to the unboosted case, as can be seen in Figure 6. We may nevertheless expect young black holes that have not yet been spun down by other astrophysical mechanisms to align along curves  of constant $c_\sigma$ in the Regge plot, so that the statistical distribution of black holes along such curves could be used to infer the presence of non-trivial extra-dimensional charges.

%%%%%%%%%%%%%%%%%%%%%%%%%%%%%%%%%%%%%%%%%%%%%%%%%%%%%%%%%%%%%%%%%%%%%%%%%%%%%%%%%%%%%%%%%%%%%%%%%%%%%%%%%%%%%%%%%%%%%%%%%%%%%%%%%%%%%
%%%%%%%%%%%%%%%%%%%%%%%%%%%%%%%%%%%%%%%%%%%%%%%%%%%%%%%%%%%%%%%%%%%%%%%%%%%%%%%%%%%%%%%%%%%%%%%%%%%%%%%%%%%%%%%%%%%%%%%%%%%%%%%%%%%%%
%%%%%%%%%%%%%%%%%%%%%%%%%%%%%%%%%%%%%%%%%%%%%%%%%%%%%%%%%%%%%%%%%%%%%%%%%%%%%%%%%%%%%%%%%%%%%%%%%%%%%%%%%%%%%%%%%%%%%%%%%%%%%%%%%%%%%

\section{Conclusion}

In this work we have analyzed the formation of unstable superradiant bound states for massive scalar fields in a boosted black string geometry in five dimensions. We have used both an analytical matching technique and a forward integration numerical procedure to compute the spectrum of such bound states as a function of the mass coupling $\mu M$ and for different values of the black string spin and boost parameters. Since Kaluza-Klein modes are generically too heavy for superradiant instabilities to develop in realistic scenarios, we have focused on the zero-modes of a bulk scalar field, with similar results for brane-bound particles. 

We have found that the boost along the compact extra-dimension, equivalent to a non-zero KK-charge and non-trivial dilaton and gauge field configurations after dimensional reduction, decreases the upper bound on the bound state frequencies for the onset of the superradiant instability, in agreement with previous analyses \cite{Dias:2006zv}. For small boosts, the states are more tightly bound to the black hole than for unboosted geometries, which consequently enhances the maximum growth rate of the superradiant instability. Above a critical boost parameter, however, this cannot compensate for the decrease in the upper limit of the superradiant spectrum, and the maximum growth rate is thus suppressed. In practice, KK-momentum may boost the black hole bomb mechanism by a factor of about 56\% for near-extremal geometries and a critical value $\cosh\sigma\simeq 2.7$.  

Given that large astrophysical black holes are most likely described by such black string or black brane geometries, our results suggest that the black hole bomb mechanism may be used not only to probe the existence of ultra-light bosonic states such as axions or hidden photons, but also to study non-trivial properties of compact extra-dimensional models, such as the KK-charge neutrality of the universe. As an example, we have determined the upper bound on the spin-mass ratio of a black string for the onset of superradiant instabilities with a light axion field, illustrating how an extra-dimensional boost leads to deeper and wider gaps in the associated Regge plot. There is also a boost-dependent upper bound on the spin-mass ratio, which is independent of the existence of unstable superradiant modes. This already poses strong constraints on a possible universal boost parameter with current observational data, although significant non-universal KK-charges are still allowed.

An important feature of our results resides in the fact that they are completely independent of the size of the compact extra-dimensions. On one hand, this implies that superradiant instabilities cannot distinguish between a higher-dimensional black string/brane at rest and a four-dimensional Kerr black hole. On the other hand, this provides a unique probe of non-trivial geometrical properties of extra-dimensional models that are independent of the KK mass scale, being complementary to dedicated collider searches, in particular at the LHC.

Although we have focused on the simplest non-trivial configuration in a model with a single extra-dimension, we hope that our results lead to further exploration of the black hole bomb effect in other geometries and in higher dimensions. This work also suggests a broader phenomenological potential for this effect, further motivating the need for precision black hole measurements and the search for additional observables, particularly those sentitive to the growth rate of the superradiant instability.

%%%%%%%%%%%%%%%%%%%%%%%%%%%%%%%%%%%%%%%%%%%%%%%%%%%%%%%%%%%%%%%%%%%%%%%%%%%%%%%%%%%%%%%%%%%%%%%%%%%%%%%%%%%%%%%%%%%%%%%%%%%%%%%%%%%%%
%%%%%%%%%%%%%%%%%%%%%%%%%%%%%%%%%%%%%%%%%%%%%%%%%%%%%%%%%%%%%%%%%%%%%%%%%%%%%%%%%%%%%%%%%%%%%%%%%%%%%%%%%%%%%%%%%%%%%%%%%%%%%%%%%%%%%
%%%%%%%%%%%%%%%%%%%%%%%%%%%%%%%%%%%%%%%%%%%%%%%%%%%%%%%%%%%%%%%%%%%%%%%%%%%%%%%%%%%%%%%%%%%%%%%%%%%%%%%%%%%%%%%%%%%%%%%%%%%%%%%%%%%%%

\acknowledgments{
I would like to thank Carlos Herdeiro, John March-Russell, Paolo Pani, Vitor Cardoso and Leonardo Gualtieri for their useful comments and suggestions. This work was supported by STFC (United Kingdom).
}

%%%%%%%%%%%%%%%%%%%%%%%%%%%%%%%%%%%%%%%%%%%%%%%%%%%%%%%%%%%%%%%%%%%%%%%%%%%%%%%%%%%%%%%%%%%%%%%%%%%%%%%%%%%%%%%%%%%%%%%%%%%%%%%%%%%%%
%%%%%%%%%%%%%%%%%%%%%%%%%%%%%%%%%%%%%%%%%%%%%%%%%%%%%%%%%%%%%%%%%%%%%%%%%%%%%%%%%%%%%%%%%%%%%%%%%%%%%%%%%%%%%%%%%%%%%%%%%%%%%%%%%%%%%
%%%%%%%%%%%%%%%%%%%%%%%%%%%%%%%%%%%%%%%%%%%%%%%%%%%%%%%%%%%%%%%%%%%%%%%%%%%%%%%%%%%%%%%%%%%%%%%%%%%%%%%%%%%%%%%%%%%%%%%%%%%%%%%%%%%%%

\end{document}